\documentclass[preprint,times]{aastex}
\usepackage{natbib}

\usepackage{graphicx}
\usepackage{epsf}

\begin{document}
\title{A Near-Infrared Search for Silicates in Jovian Trojan Asteroids }
\author{Bin Yang$^{1}$ and David Jewitt$^{2,}$$^{3}$
\affil{$^1$ Institute for Astronomy, University of Hawaii, Honolulu, HI
96822 \\ 
$^2$ Department of Earth and Space Sciences, Institute for Geophysics and Planetary Physics,University of California, Los Angeles, CA 90095 \\
$^3$ Department of Physics and Astronomy, University of California, Los Angeles, CA 90095} 
\email{yangbin@ifa.hawaii.edu, jewitt@ucla.edu}}

\begin{abstract} 
We obtained near-infrared (0.8 - 2.5 $\mu$m) spectra of seven Jovian Trojan asteroids that have been formerly reported to show silicate-like absorption features near 1-$\mu$m. Our sample includes the Trojan (1172) Aneas, which is one of three Trojans known to possess a comet-like 10-$\mu$m emission feature, indicative of fine-grained silicates. Our observations show that all seven Trojans appear featureless in high signal-to-noise ratio spectra. The simultaneous absence of the 1-$\mu$m band and presence of the 10 $\mu$m emission can be understood if the silicates on (1172) Aneas are iron-poor. \\

In addition, we present  near infrared observations of five optically grey Trojans, including three objects from the collisionally produced Eurybates family. The five grey Trojans appear featureless in the near infrared with no diagnostic absorption features. The near infrared spectrum of Eurybates can be best fitted with the spectrum of a CM2 carbonaceous chondrite, which hints that the C-type Eurybates family members may have experienced aqueous alteration. \end{abstract}


\section{Introduction}

The Jovian Trojan asteroids inhabit two large co-orbital swarms at the L4 and L5 Lagrangian points of the Sun-Jupiter system. Their unique location, in between the largely rocky asteroids of the main-belt and the icy Kuiper belt objects beyond Neptune, suggests that the Trojans may  preserve important distance- and temperature-dependent compositional trends.  It has long been recognized that the Trojans are observationally similar to the nuclei of comets, in terms of their slightly reddish colors \citep{jewitt:1990} and their low albedos \citep{fernandez:2003}. They are quite distinct from, for example, the S-type asteroids that dominate the inner regions of the main asteroid belt \citep{gradie:1982}. On the other hand, while comet nuclei are known to be ice-rich, no water ice has ever been identified on a Trojan asteroid despite the expenditure of considerable observational effort \citep{jon90,luu94,dum98,emery:2003,yang:2007,fornasier:2007}.  Water ice is expected to be an important constituent of bodies formed beyond the snow line, where water is thermodynamically stable in the solid phase.  The absence of surface water ice may be due, as on the nuclei of comets, to the formation of a refractory surface mantle that is a by-product of past sublimation \citep{jewitt:2002}. If so, the surface properties of the Trojans reveal the nature of the refractory components of these bodies, with ice perhaps remaining stable only at depths to which optical and infrared photons cannot penetrate.

The source regions from which the Trojans were captured remain unknown. Published hypotheses fall into two main categories: one suggests that Trojans are planetesimals captured from the vicinity of the orbit of the giant planet \citep{marzari:1998,fleming:2000,chiang:2005}, while the other suggests that the Trojans might have formed in the Kuiper belt at an early epoch, and then were scattered and captured into their current orbits \citep{morbidelli:2005}. Trojans accreted near Jupiter's orbit (5 AU and temperatures $\sim$100K to 150 K) and in the much more distant Kuiper belt (30 AU and 50 K) should differ compositionally. Therefore, the chemical composition of the Trojans should set important constraints on the plausible source regions of these objects. 

Recently, the thermal emissivity spectra of three Trojan asteroids, namely (624) Hektor, (911) Agamemnon and (1172) Aneas, were reported by \cite{emery:2006}. These three Trojan spectra, taken by the Spitzer Space telescope, show an almost identical emissivity maximum near 10-$\mu$m with two weaker features near 20 $\mu$m. These features are difficult to match with laboratory meteorite spectra or synthetic mineral spectra but closely resemble the 10 $\mu$m and 20 $\mu$m features found in the spectra of comets \citep{emery:2006}. The cometary emission features are attributed to the Si-O vibrational modes in small silicate grains \citep{ney:1977,crovisier:1997,wooden:1999,stansberry:2004}. Silicate grains with radii larger than a few tens of microns do not exhibit 10 and 20 $\mu$m spectral features but display black body-like spectra because large grains are optically thick at these wavelengths \citep{rose:1979}. The detection of the thermal emission features suggests a very porous distribution of small silicate grains on the surfaces of these Trojans. This finding is consistent with previous studies \citep{cruikshank:2001,emery:2004}, which demonstrated that synthetic models incorporating amorphous carbon and amorphous pyroxene can adequately reproduce the common red slopes of the Trojan spectra in the near infrared (NIR). The porous distribution of small particles might have been caused by past or on-going cometary activity. However, the process is not understood in detail.

Howell (1995) performed visible and infrared spectrophotometry of 17 Trojans and reported that nine out of seventeen Trojans showed an absorption feature near 1.0 $\mu$m, when the data from the Eight-Color Asteroid Surve \citep{zellner:1985} were included. If real, this feature would be consistent with the diagnostic 1-$\mu$m silicate band, which is attributed to Fe$^{2+}$ crystal field transitions in iron-bearing silicates \citep{burns:1981, king:1987, sunshine:1998}. The detection of the 1-$\mu$m feature in the NIR and the strong 10-$\mu$m features in the mid-infrared (MIR) suggest that the surfaces of Trojan asteroids could be dominated by silicate-rich mantles. However, the detections by Howell (1995) were based on multi-color photometry taken with single-element detectors and remain observationally unconfirmed. Accordingly, the first objective of this paper is to verify the presence of the 1-$\mu$m absorption feature at higher spectral resolution and signal-to-noise-ratios (SNRs). 

Most, but not all, of the Trojans with taxonomical classifications are P- or D-types based on their optical reflectance spectra. The latter are generally featureless with a moderate or very steep red spectral slope longward of 0.55 $\mu$m \citep{tholen:1989, bus:2002}. A common belief among planetary scientists is that Trojans are exclusively red. But is this really true? Recently, several Trojans were found to have neutral or grey colors in the optical \citep{bendjoya:2004, dotto:2006, szab:2007}. Many of these grey Trojans belong to one dynamical family, the Eurybates family \citep{beauge:2001}. Members of this family were reported to have optical spectra with normalized spectral slopes, $S'$, of no more than a few percent per 1000\AA, which closely resemble the C-type asteroid spectra \citep{dotto:2006, fornasier:2007}. 

Given the dominating fraction of the P- and D-types among the Trojan population, the rareness of the C-type-like grey Trojans in itself is interesting. More importantly, \cite{emery:2009} reported that the red-sloped Trojans generally show stronger thermal emission features than those that have neutral (grey) spectra. However, Trojans that have been studied in the NIR are mainly (red sloped) D-types and the NIR properties of C-type Trojans remain largely under-sampled. Therefore, the second objective of this paper is to investigate surface properties of grey Trojans via NIR spectroscopy to improve our knowledge of these objects. We observed three Eurybates family members in the NIR, together with two non-family grey objects to investigate: a) whether the Eurybates family possesses any special traits, b) whether grey objects with neutral optical spectra display any diagnostic features that are indicative of their mineralogy.

\section{Observations and Data Reduction}
The NIR spectra were obtained using the NASA Infrared Telescope Facility (IRTF) 3-m telescope atop Mauna Kea, Hawaii. A medium-resolution 0.8-5.5 $\mu$m spectrograph (SpeX) was used, equipped with a Raytheon 1024 $\times$1024 InSb array having a spatial scale of 0.$\!\!^{\prime\prime}$15 pixel$^{-1}$ \citep{rayner:2003}. The low-resolution prism mode was used to cover an overall wavelength range from 0.8 $\mu$m to 2.5 $\mu$m for all of our observations. We used a 0.8$^{\prime\prime}$x15$^{\prime\prime}$  slit that provided an average spectral resolving power of $\sim$130. To correct for strong telluric absorption features from atmospheric oxygen and water vapor, we used G2V stars that were close to the scientific target both in time and sky position as telluric calibration standard stars. The G2V stars generally have spectral slopes similar to the Sun and they also served as solar analogs for computing relative reflectance spectra of the scientific targets. Usually, two standard stars (with airmasses different from the target and from each other by $\leqslant$ 0.1) were observed before and after each scientific object to ensure good telluric calibration and to reduce systematic errors introduced into the reflectance spectra due to potentially imperfect solar analogs. During our observations, the slit was always oriented along the parallactic angle to minimize effects from differential atmospheric refraction. 

The SpeX data were reduced using the SpeXtool reduction pipeline \citep{Cushing:2004lq}. All frames were flatfielded. Bad pixels were identified and removed using flat field frames as well as a stored bad pixel mask. The first-order sky removal was achieved via subtracting image pairs with the object dithered along the slit by an offset of 7.$\!\!^{\prime\prime}$5. Wavelength calibration was accomplished using argon lines from an internal calibration lamp. The reflectivity spectra, $S(\lambda)$,  were computed by dividing the calibrated Trojan spectra by the spectra of nearby Solar analogs.  

A journal of observations is provided  in Table 1.

\section{Result}
\subsection{The 1-$\mu$m Silicate Absorption Feature}
\begin{figure}
\begin{center}
\includegraphics[width=5.5in, angle=90]{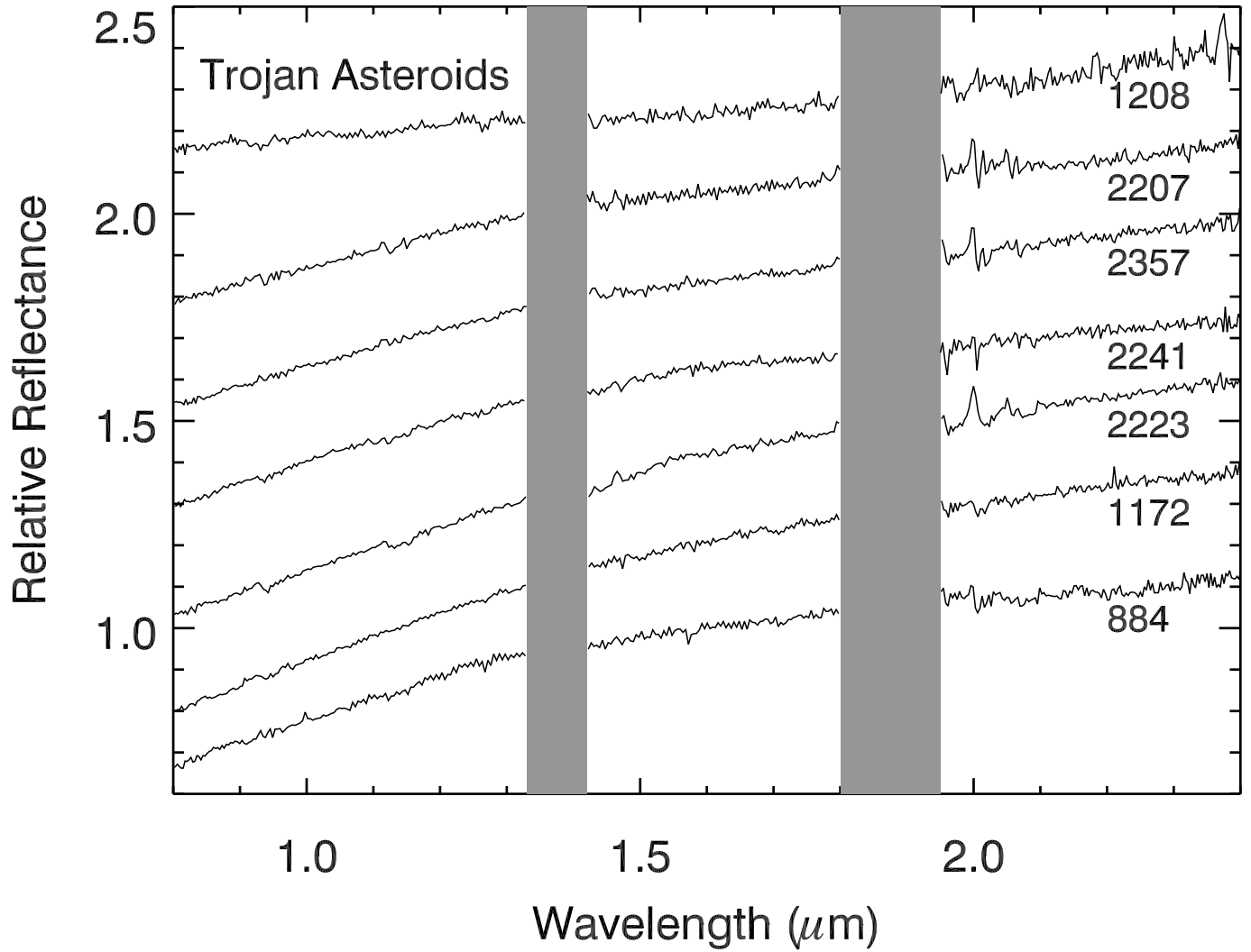}
\caption{Near-infrared spectra of seven Trojans that have been previously reported to show absorption features near 1.0 $\mu$m. The reflectance spectra are normalized at 1.7 $\mu$m and have been vertically offset for clarity. }
\label{f1}
\end{center}
\end{figure}

\begin{figure}
\begin{center}
\includegraphics[width=4.5in]{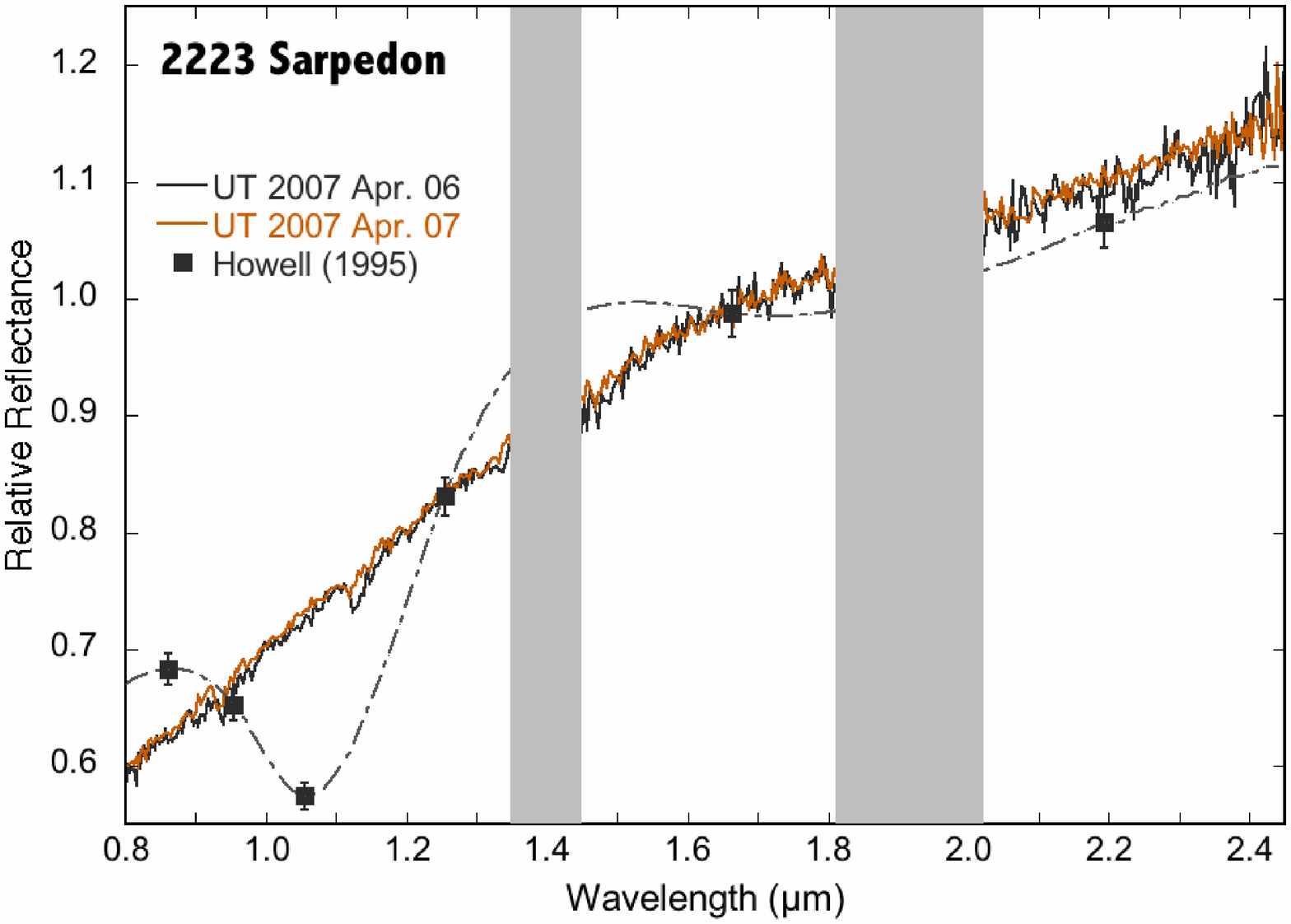}
\includegraphics[width=4.5in]{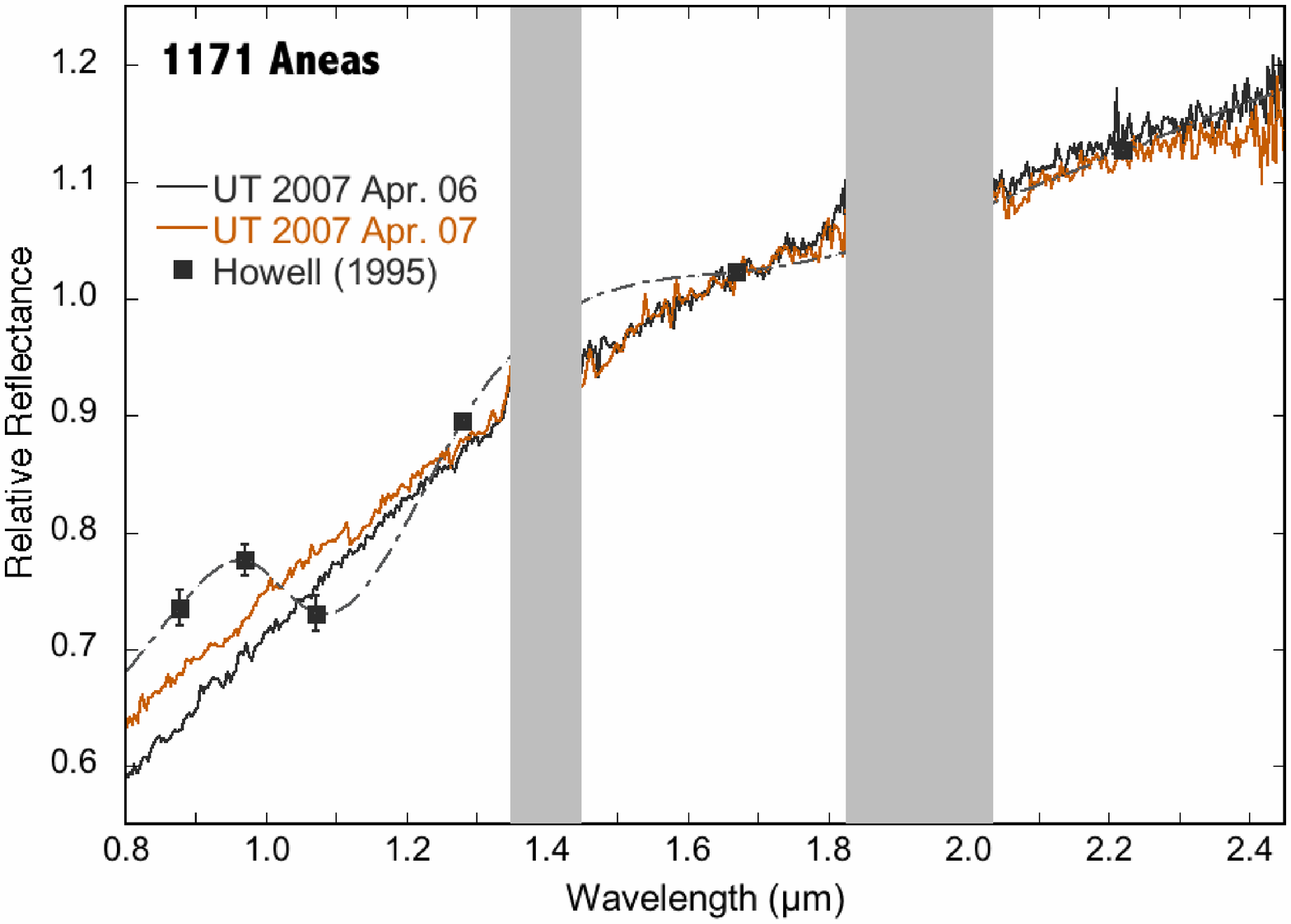}
\caption{Photometric observations from \cite{howell:1995} compared with our newly obtained spectra of (1172) Aneas and (2223) Sarpdon. The black and orange solid lines represent our observations obtained on UT 2007 Apr.\ 06 and Apr.\ 07 respectively. The filled squares are observations presented in \cite{howell:1995}. All the spectra are normalized at 1.7 $\mu$m. Dashed lines are cubic spline fits to the Howell data, added to guide the eye.}
\label{f2}
\end{center}
\end{figure}

We observed seven Trojans that were previously reported to display a distinctive absorption feature near 1.0 $\mu$m (Howell 1995). Our newly obtained NIR spectra are illustrated in Figure \ref{f1}. All seven Trojans appear featureless in our data and have reddish spectral slopes from 0.8 to 2.5 $\mu$m. Detailed comparisons between our observations and the ones of Howell (1995) are presented in Figure \ref{f2} for (2223) Sarpedon and (1172) Aneas. The former has the strongest 1-$\mu$m absorption band among all seven Trojans in Howell (1995), while the latter was observed using NASA's Spitzer space telescope and found to exhibit a significant 10-$\mu$m emission feature. We normalized both our data and the photometric data from (Howell 1995) at 1.7 $\mu$m. In Figure \ref{f2}, the black and the orange solid lines represent the NIR spectra of Sarpedon and Aneas that were taken on UT 2007 Apr. 06 and 07, respectively. The filled black squares are photometric measurements of these two objects taken from Howell (1995). The dashed lines are cubic spline fits to the photometric points. Our spline fitting curves are similar to the original fits in Howell (1995).  We found that our data are broadly consistent with Howell's measurements in the wavelength range from 1.0 to 2.5 $\mu$m, except for the deviation near 1.1 $\mu$m, where Howell's data show a strong absorption band while we did not see this feature in our data. 

\begin{figure}
\begin{center}
\includegraphics[width=6.5in]{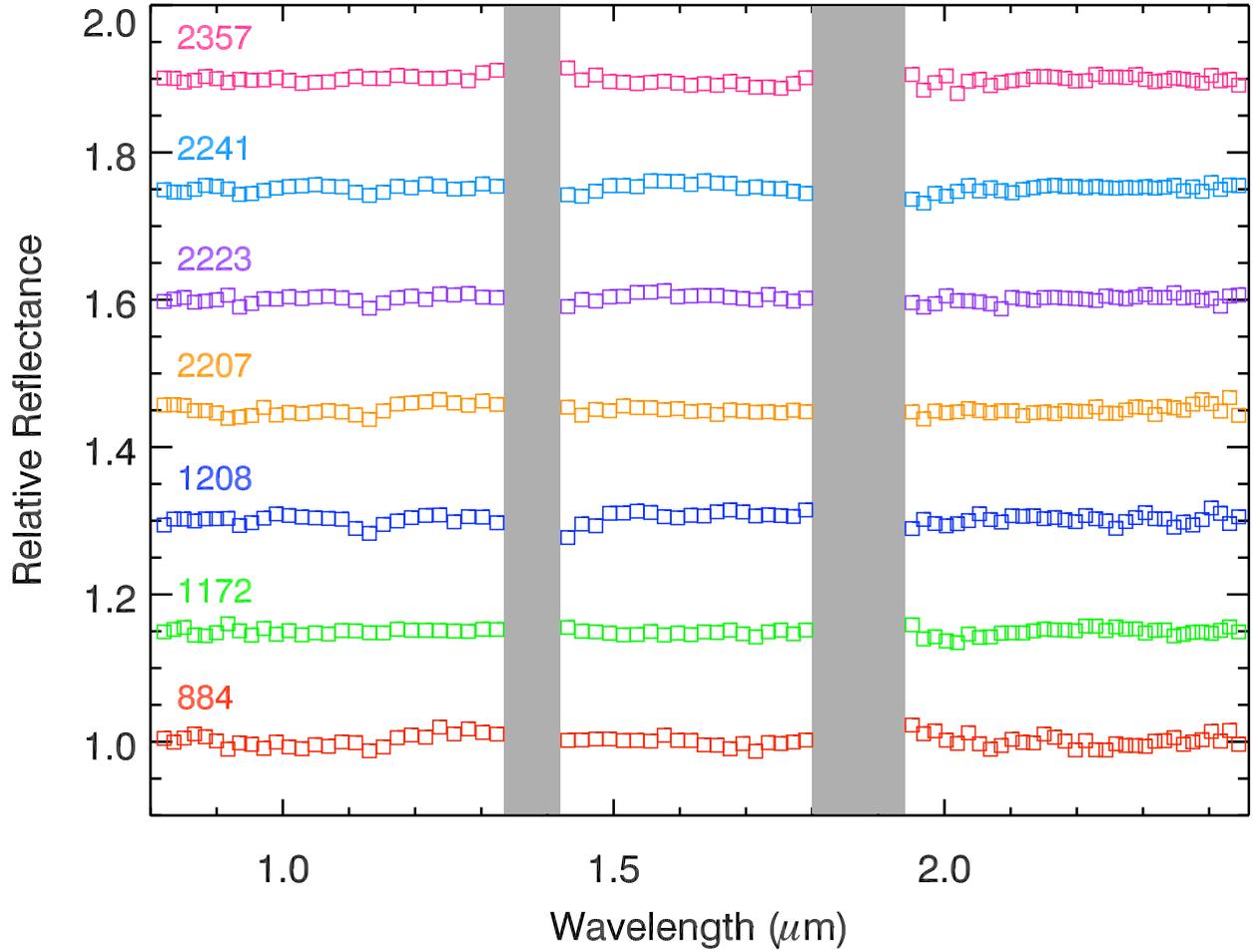}
\caption{Residuals in slope-removed spectra of Trojans to enhance possible absorption features near 1.0 $\mu$m. The spectra have been vertically offset for clarity. No significant features are present.  The minor absorptions near 1.1, 1.4 and 2.0 $\mu$m are due to the incomplete telluric calibration. No silicate absorption features are detected at 1.0 and 2.0 $\mu$m. }
\label{f3}
\end{center}
\end{figure}

To search for possible weak features, we examined the continuum-removed residual spectra, shown in Figure \ref{f3}. As illustrated in Figure \ref{f1}, the continua of several Trojans in our sample are non-linear and show a turnover near 1.5 $\mu$m. Therefore, we split each spectrum into two parts then fit and subtract the continuum from each part separately. At short wavelengths (0.8 - 1.5 $\mu$m), a linear regression model was used to fit the continua. At longer wavelengths (1.5 - 2.5 $\mu$m), a second-order polynomial function was used. Note that we excluded regions (1.35 -1.45 $\mu$m, 1.8 -1.95 $\mu$m) that are contaminated by the telluric absorptions from continuum fitting. Some small variations present near 0.9, 1.4, 2.0 and 2.1 $\mu$m are caused by imperfect telluric calibration. Based on the noise level in the SpeX data, we conclude that no absorption features deeper than 1\% were present in the seven observed Trojans. The data from Howell (1995) were also used to estimate band depths and the results are compared in Table \ref{abs_lim}. Given that our data were taken at far higher spectral resolution and signal-to-noise ratios, we conclude that the previously reported 1-$\mu$m band is unlikely to be real. The possibility that we might have missed the 1-$\mu$m feature because of azimuthal non-uniformities on the rotating Trojans seems unlikely, both because we would have to be unlucky with all seven objects and because we observed each on two different nights. Furthermore, observations of other asteroids generally show that azimuthal spectral variations are either small or undetectable.

\subsection{Grey Trojans}
As shown in Figure \ref{f1}, the spectrum of (1208) Troilus is distinct from the other six Trojans. Firstly, Troilus exhibits a nearly flat near-infrared spectrum with a spectral slope $S'$ = $dS/d\lambda \sim$ 1.2\%/10$^{3}$\AA~(obtained via fitting a linear function to the entire spectrum from 0.8 to 2.5 $\mu$m). In contrast, the other Trojans have steeper slopes ($S' >$ 3.0\%/10$^{3}$\AA) in the same spectral region. Secondly, the spectrum of Troilus has a nearly unchanging slope with wavelength ($S''$ = $d^2S/d\lambda^2 \sim$ 0), whereas the other Trojans show a flattening ($S'' <$ 0), with a turnover near 1.5 $\mu$m. Optical observations show that Troilus has a grey spectrum ($S' \sim$ 4.8\%/10$^{3}$\AA) from 4000 to 9000 \AA~where the other six Trojans have exclusively red-sloped ($S' >$ 8\%/10$^{3}$\AA) spectra \citep{jewitt:1990, bendjoya:2004}. Could (1208) Troilus, with a distinct NIR spectrum, be a special case? Or is Troilus representative of the under-studied grey Trojan group that shows neutral/grey spectra in the optical and NIR?

\begin{figure}
\begin{center}
\includegraphics[width=6in]{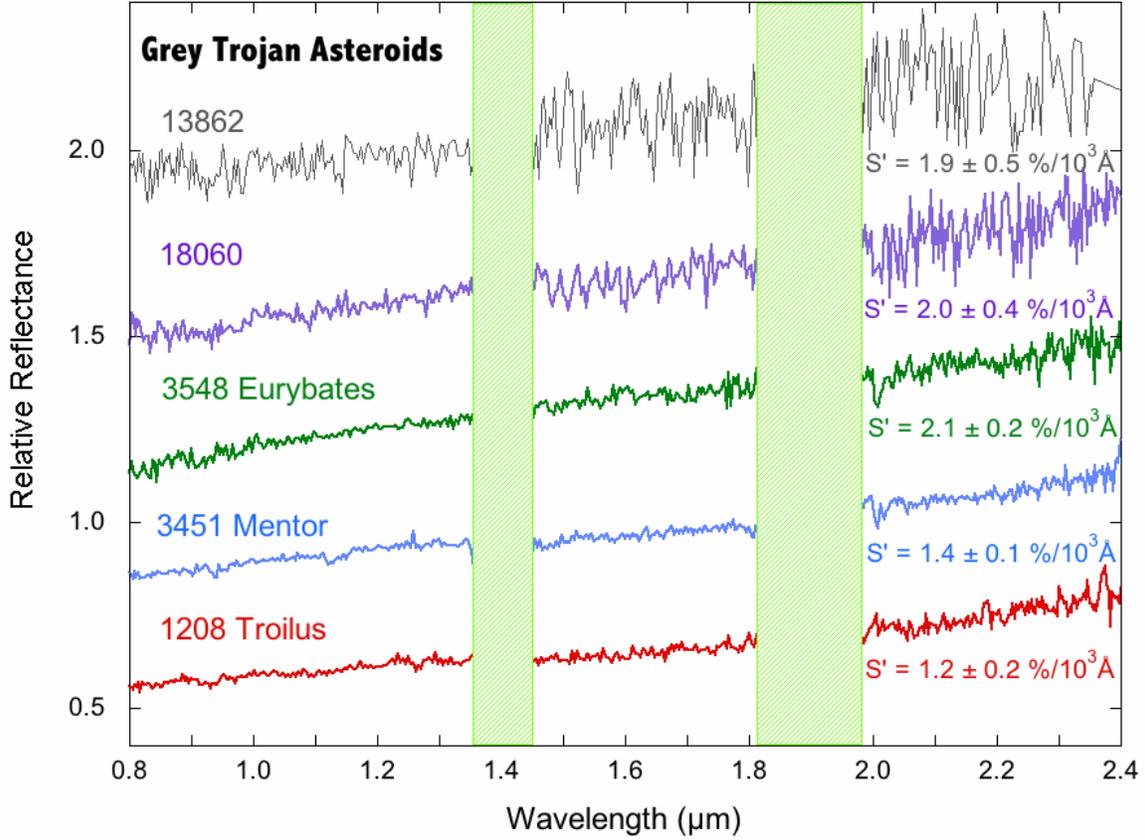}
\caption{Near-infrared spectra of Trojans that have nearly neutral optical spectral slopes. The optically neutral Trojans exhibit similar spectral slopes in the NIR and show no diagnostic spectral features. Note the especially close similarities between the Eurybates family members (e.g. 3548, 18060, 13862), all with spectral slope of $\sim$ 2\%/10$^{3}$\AA \ in the wavelength range 0.8 to 2.4 $\mu$m. The reflectance spectra are normalized at 1.7 $\mu$m and have been vertically offset for clarity. } 
\label{f4}
\end{center}
\end{figure}

To address these questions, we obtained NIR spectra of four additional optically grey Trojans in April and August, 2007 (Table 1), shown in Figure \ref{f4}. Three out of five of these grey objects are members of the Eurybates family, namely (3548) Eurybates, 18060 and 13862. Trojan (1208) Troilus and (3451) Mentor are non-Eurybates grey objects in our sample. The five optically grey Trojans exhibit very similar NIR spectra, with $S' \sim$ (1 to 2) \%/1000\AA \ and $S'' \sim$ 0 in the 0.8 $\le \lambda \le$ 2.4 $\mu$m range. Figure \ref{f4} shows that (1208) Troilus is a common grey Trojan, in spite of having an inclination ($i$ = 33.5$^{\circ}$), inconsistent with that of the Eurypates family ($\bar{i}$ = 7.2$^{\circ}$). We note that the three Eurybates family members show essentially identical NIR spectra, within the uncertainties of measurement. 

Since the spectral properties of the three Eurybates Trojans closely resemble each other, we considered only the spectrum of (3548) Eurybates (with the highest SNR) for further analyses. The IRAS albedo of 3548 is 0.054 $\pm$ 0.007 \citep{tedesco:2002}, which is consistent with the common albedos of the main belt C-type asteroids that generally have neutral spectra and low albedos in the optical. The C-types are thought to be composed of hydrated silicates, carbon, and organic compounds \citep{gaffey:1993}. In particular, it has been suggested that C-type asteroids may be parent bodies for CI and CM carbonaceous chondrites \citep{bell:1989,gaffey:1993}. Accordingly, we compared the spectrum of 3548,  in the wavelength range from 0.8 to 2.45 $\mu$m, with the laboratory spectra of common carbonaceous meteorites from the Brown University Keck/NASA Relab Spectra Catalog. We mainly compared our data with the CI and CM chondrite samples, because our grey Trojan samples and the C-type asteroids are spectrally alike in the optical. For completeness, we also examined the similarity between (3548) Eurybates and a few other carbonaceous chondrites, such as CO and CV meteorites. We further compared our spectra with that of the Tagish Lake meteorite, which has been independently proposed as a spectral analog of the D-type asteroids \citep{hiroi:2001}. 

\begin{figure}
\begin{center}
\includegraphics[width=6in]{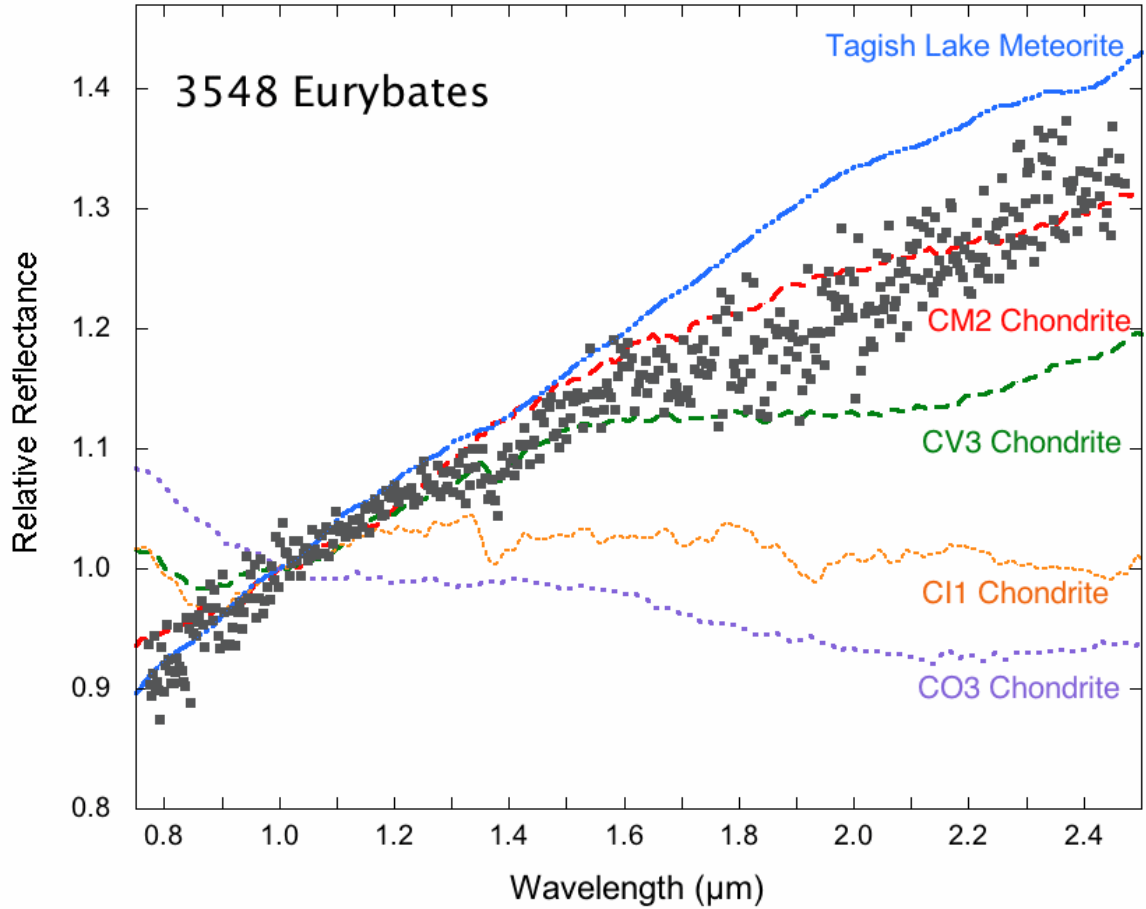}
\caption{Comparison between the spectrum of (3548) Eurybates and common carbonaceous chondrites, including the famous Tagish Lake meteorite. The Trojan spectrum (shown as solid squares) and the meteorite spectra are all normalized at 1.0 $\mu$m. Like main-belt C-type asteroids, the NIR spectrum of (3548) Eurybates closely resembles that of the CM2 meteorite.}
\label{f5}
\end{center}
\end{figure}

We searched for spectral analogs to (3548) Eurybates among all the available CI and CM chondrites in the RELAB database using a $\chi^{2}$-test. Our results are shown in Figure \ref{f5} with all the spectra normalized at $\lambda = 1.0 ~\mu$m. The spectrum of a laser-irradiated CM2 meteorite (Mighei), shown in red, matches the Trojan spectrum the best, except for minor discrepancies at 1.4 and 1.9 $\mu$m which are due to incomplete telluric line removal. The Mighei samples were sieved to several partice size fractions and a ND-YAG multiple-pulse laser was applied under a vacuum of 10$^{-4}$ Hg \citep{moroz:2004b,shingareva:2004}. The reflectance spectra of original Mighei samples have neutral spectral slopes and exhibit a major absorption band at 2.7 $\mu$m and a weak absorption feature at 3.4 $\mu$m, which are due to phyllosilicates and hydrocarbons respectively. After irradiation, the spectra of processed Mighei are significantly reddened and the absorption features are greatly weakened. We found that the spectrum of irradiated Mighei coarse powders (bkr1ma062 , grain size $>$ 200 $\mu$m) fits the spectrum of Eurybates the best. Mighei is a typical CM2 and its optical albedo was measured to be 0.043 by \cite{gaffey:1976}. The comparable optical albedos and the similar spectra in the visible and NIR are consistent with the idea that grey Trojans are compositionally like the CM2 meteorites, being composed of fine-grained phyllosilicate (clay) minerals, diamond, silicon carbide, and graphite \citep{huss:2003}. However, we need to point out that the spectral similarity between a featureless astronomical spectrum and a laboratory meteorite spectrum, no matter how suggestive, can never provide definitive proof of the actual mineralogy,  given the absence of diagnostic absorption features \citep{clo90}. 

\section{Discussion}
\subsection{Missing Silicate Absorption}
\citet{emery:2006} thoroughly searched for spectral analogs for the Trojan emissivity spectra amongst meteorites, minerals and small body spectral databases. They found that neither the thermal IR spectra of the Tagish Lake meteorite nor of granular mixtures of amorphous silicates with amorphous carbon can reproduce the observed 10 $\mu$m plateau along with the minor features in the 15 - 30 $\mu$m region. This is despite the success of these models in the visible and the NIR (0.4 $< \lambda < $ 2.5 $\mu$m) \citep{hiroi:2001, cruikshank:2001, emery:2004}. Surprisingly, the mid-infrared (MIR) spectra of comets, namely Hale-Bopp and Schwassmann-Wachmann 1 (SW1), provide the best match to the Trojan thermal spectra, in terms of the overall profile and the alignments of the centers of the emission bands. 

In contrast to the short history of the thermal observations of asteroids, the 10 $\mu$m IR emission feature observed in astronomical spectra has been a subject of study for more than 30 years. This feature is believed to be associated with the Si-O stretching mode in silicate grains \citep{ney:1977} and the minor features between 16 and 35 $\mu$m are due to the bending mode of the Si-O bond \citep{wooden:1999}. Numerous investigations have been carried out to characterize the physical properties, such as the mineralogy, crystallinity and grain sizes of silicate particles via fitting synthetic models to these emission features \citep{stephens:1979,mukai:1990, gehrz:1992, bradley:1992,kimura:2008}. The band centers and shapes of the silicate emission features are diagnostic of the mineralogical compositions \citep{Hanner:1999}. However, the strengths of the emission bands depend primarily upon the temperature of the silicate grains and the temperature, in turn, is controlled by the grain size, the Mg/Fe ratio and the total amount of incorporated dark materials \citep{Hanner:1999}. As an example, high quality thermal infrared spectra of comet Hale-Bopp show that this comet consists of multiple silicate components, such as crystalline pyroxene (enstatite), olivine (forsterite) and amorphous silicate groups, with a large fraction of sub-micron grains \citep{hayward:1997, wooden:1999, harker:2002}. 

The radiative properties of comets and asteroids must be compared with caution, because multiple scattering is likely to play a more important role in shaping the emissivity spectra from a (comparatively dense) regolith than from a coma \citep{moersch:1995, emery:2006}. Cometary spectra are generated from well-separated particles in an optically-thin coma, although multiple scattering may be important inside ``dust-ball'' particles, consisting of loose aggregates of tiny constituent grains (Kimura et al. 2006). The exceptionally good match at 10$\mu$m between the Trojan and cometary spectra appears significant, given that a large number of samples have been considered by \cite{emery:2006}.  Presumably, the Trojan regolith supports a very low density upper layer, in which small silicate particles can be heated by the Sun without strong radiative coupling to the denser regolith beneath. Fine-grained silicate mixtures (olivines, pyroxenes, phyllosilicates and carbonates) can generally reproduce the major features in the Trojan spectra, including a blunt emission peak near 10 $\mu$m and a moderate enhancement in emissivity in the region from 18 to 25 $\mu$m, although some discrepancy was observed when the minor features were taken into account \cite{emery:2006}. While olivine and pyroxene are consistent with the 10 $\mu$m emission spectrum, these materials also possess absorption features in the NIR, including the characteristic 1-$\mu$m band. The MIR spectra of (1172) Aneas and (911) Agamemnon both exhibit  emission features at 10 $\mu$m but no 1-$\mu$m band was observed in the NIR on these objects (this paper and \cite{yang:2007}). 

Olivine and pyroxene are both ferromagnesian minerals with formula [(Mg$_{x}$Fe$_{1-x})$SiO$_{4}$] and [(Mg$_{x}$Fe$_{1-x})$SiO$_{3}$] respectively, where $x$ represents the magnesium abundance.  As mentioned earlier, the broad absorption band near 1.0 $\mu$m is attributed to a crystal field absorption of Fe$^{2+}$ in the silicate structure. Thus, the strength of the 1-$\mu$m absorption band directly depends on the iron abundance in the host silicates, all else being equal. As such, iron-poor silicates have relatively weak 1-$\mu$m absorption features. In one extreme case, pure enstatite (Mg$_{2}$Si$_{2}$O$_{6}$), as an iron-free silicate, appears featureless in the NIR with no trace of an absorption band near 1.0 $\mu$m \citep{hardersen:2005}. Consistently, previous studies of comets have revealed that the observed silicate emission features in the cometary MIR spectra are dominated by Mg-rich silicate minerals \citep{crovisier:1997, wooden:1999, Harker:2007}. Moreover, the best fitting silicate model in \cite{emery:2006} consists mainly iron-poor (or magnesium-rich) silicates. As such, it is possible that the simultaneous absence of the 1 $\mu$m silicate absorption and the presence of the 10-$\mu$m silicate emission is due to the silicate components of those Trojans being iron-poor.  

An immediate problem with the iron-poor interpretation is that low-iron silicates have relatively high albedos \citep{lucey:2008}. For example, (44) Nysa is believed to be an iron-depleted asteroid \citep{zellner:1975} and the measured geometric albedo of this object is 0.546 $\pm$ 0.067 \citep{tedesco:2002}. In contrast, Trojans are very dark objects with an average albedo of $\bar{p}_{v}$= 0.041 $\pm$ 0.002 \citep{fernandez:2003}. Carbonaceous materials mixed with low-iron silicates might significantly suppress the albedo. However, both carbonaceous materials and pure silicates are spectrally neutral and they are not sufficient to account for the red slopes of Trojan spectra. Previous studies of Trojans \citep{cruikshank:2001, emery:2004} demonstrated that oxidization of ferrous iron carried in amorphous pyroxene can greatly redden the visible and NIR spectra of this material. Consequently, ferrous-iron bearing silicates were used as the main component by these authors in their models to simulate the Trojan spectra. However, if the silicates on Trojans are iron-poor, as suggested by the MIR observations and the absence of the 1-$\mu$m band, then the redness of Trojan spectra would require other explanations. 

The origin of the Trojan's red color should not be considered in isolation. We must frame this color problem in the context of the whole solar system. A systematic variation of asteroid spectral type with respect to heliocentric distance has been noticed by \cite{gradie:1982} with the S-complex mainly located in the inner belt, the C-complex dominating the mid-belt around 3 AU and the featureless red D-types mainly found beyond 3.5 AU. This original finding is supported by numerous subsequent observations \citep{ivezic:2001, bus:2002}. In addition, the detection of ultra-red materials on the distant Kuiper Belt objects \citep{jewitt:2002} further supports this correlation. This well-established radial distribution of patterned taxonomical types suggests that the compositions of different small body populations are strongly influenced by their distances to the Sun, in other words, by their temperatures. It has been proposed that red spectral slopes may be due to a long wavelength wing of a broad UV feature caused by absorption in polycyclic aromatic hydrocarbons \citep{moroz:1998}. Trojans probably formed beyond 5 AU, where temperatures are low enough for dark and red organic matter to be thermodynamically stable. \cite{emery:2004} pointed out that the absence of absorptions in the 3-4 $\mu$m wavelength region limits the possible abundance and species of organics on Trojans. However, the quality of the existing 3-micron observations of Trojans asteroids is not sufficient to rule out the presence of possible organic materials with weaker absorption bands in the 3 $\mu$m region. Besides organic materials, the carbon-rich (4 to 5 wt.\%) Tagish Lake meteorite exhibits a Trojan-like red spectrum in the visible and NIR \citep{brown:2000, hiroi:2001} suggesting (but not proving) a
carbon-red color connection. No absorption band was detected in the reflectance spectrum of the Tagish Lake meteorite in the 3-4 $\mu$m
region, although an absorption band of OH bond in phyllosilicates was observed at 2.7 $\mu$m \citep{hiroi:2001}, a wavelength where few useful ground-based data exist. Therefore, Trojans may be composed of iron-poor silicates and carbonaceous compounds. 

We note that the iron-poor silicates scenario is able to explain the non-detection of the 1 $\mu$m absorption band but it is not a unique explanation. For example, many laboratory investigators have documented that particle size is a crucial parameter that greatly affects reflectance spectra, especially the strength of absorption features \citep{hunt:1968,salisbury:1992,mustard:1997}.  According to \cite{mustard:1997}, the 1 $\mu$m absorption feature generated by micron-sized particles has a depth $\sim$10\% of the continuum. Such a weak band is susceptible to masking by carbonaceous materials. Another possibility is that the crystallinity of silicate grains can also affect the strength of the 1 $\mu$m absorption band \citep{emery:2004}. 

\subsection{Grey Trojans and Aqueous Alteration}
The distribution of the optical colors of asteroids beyond 3.3 AU shows that the fraction of C-type objects drops rapidly as the heliocentric increases \citep{bus:2002, gil:2008, roig:2008}. Infrared observations of the main-belt asteroids reveal that many C-type asteroids exhibit a strong absorption band near 2.9-$\mu$m that is the diagnostic feature of hydrated silicates, such as phyllosilicates \citep{aines:1984, rivkin:2002}. As such, many C-type asteroids are thought to have been aqueously alterated, meaning that the chemistry and mineralogy of the host asteroid have been modified by reactions with liquid water \citep{brearley:2006}. It has been suggested that the relative lack of C-type asteroids in the Hilda and Trojan populations is a result of their large heliocentric distances and low equilibrium temperatures, which prevented water ice from melting \citep{bell:1989}. 

Recent studies of Trojan dynamical families found that several members in the ``Eurybates family'' and the ``1986WD family'' show C-type like visible spectra. Fornasier et al. (2007) noted that some C-type Eurybates Trojans show a decreased reflectance at wavelengths $\lambda < 5000$ \AA. This decline of reflectance in the blue and possibly the ultraviolet (UV) regions is similar to the UV absorption (also known as the ``UV drop-off'') that has been widely observed in the spectra of other low-albedo asteroids, and which may be an indicator of past aqueous alteration  \citep{burns:1981, feierberg:1985}. Supporting this (admittedly non-unique) interpretation, we found that the best analog to the NIR spectra of the Eurybates family members is given by a CM2 carbonaceous chondrite (Figure 5), an object which is rich in hydrated minerals. In contrast, the observations of the five grey Trojans in this study and the observations of seven Eurybates family members in \cite{deluise:2010} found no evidence of aqueous altered materials. However, given that the strong fundamental absorption bands of water ice and hydrated silicates occur beyond 2.5 $\mu$m, it is not surprising that current observations of the grey Trojans, from 0.5  to 2.5 $\mu$m, found no evidence of aqueously altered materials. 

Changes in spectral slopes due to the so-called space weathering effect have been inferred in near-earth \citep{binzel:2004} and main-belt asteroids \citep{chapman:1996, nesvorny:2005, lazzarin:2006}. However, the magnitude and even the sense of the effect depend on many unknown factors, notably the original surface composition.  For example, experiments by \cite{moroz:2004} show that bombarding complex organic materials with low energy ions (representing the solar wind) can neutralize the spectral slopes in the visible and NIR wavelength ranges. This reduction of spectral slope may be due to a progressive carbonization of the surface materials \cite{moroz:2004}. In contrast,  \cite{brunetto:2006} irradiated frozen (16-80 K) methanol (CH$_{3}$OH), methane (CH$_{4}$), and benzene (C$_{6}$H$_{6}$) with ions and found the opposite effect, with the surfaces  becoming reddened and darkened due to the formation of an organic (C-rich) refractory mantle. These opposite reddening trends would lead to opposite conclusions, if applied to the case of the Trojans.  Work by \cite{moroz:2004} would suggest that the spectrally neutral objects are old and rich in carbonaceous compounds, consistent with the evident similarity between the Eurybates members and the carbonaceous chondrites. On the contrary, the experiments of \cite{brunetto:2006} would lead to the interpretation that the Eurybates family is young and unweathered. The observation that the Eurybates family forms a tight cluster within the Menelaus clan \citep{roig:2008} suggests a younger age for the Eurybates members,  consistent with the latter interpretation. 

Our understanding of the effects of space weathering remains very limited. More laboratory studies of space weathering on icy or organic-rich materials are needed. More importantly, high quality observations of Trojan asteroids in the IR wavelength regions (especially near 3-$\mu$m and 10-$\mu$m, where the diagnostic absorption features lie), will be key to understanding the nature of Trojan asteroids. If future observations of the C-type Trojans at longer wavelengths reveal hydrated minerals, this would set a strong constraint on the chemical and thermal evolution of the Trojan population, because liquid water can only exist under particular conditions of temperature and pressure. 

\section{Summary}
We obtained new, high quality near-infrared reflection spectra of eleven Jovian Trojan asteroids, with the following results:

\begin{itemize}
   \item  No 1-$\mu$m silicate absorption feature (deeper than about 1\% of the local continuum) was found, in contradiction to previously reported spectra showing this band in these objects  \citep{howell:1995}.
  \item One of the seven objects (1172 Aneas) does show a thermal emission spectrum consistent with silicate particles (Emery et al. 2006).  The simultaneous presence of a silicate feature at 10 $\mu$m and absence of one at 1 $\mu$m can be understood if the silicates are iron-poor. But, if so, the red color of the Trojans is more likely to be due to organics than to oxidized iron (opposite to the conclusion of Cruikshank et al. 2001).
 \item The near infrared spectra of the optically neutral Trojans (the Eurybates family members) are consistent with that of a CM2 carbonaceous chondrite, which hints that these C-type Trojans may contain aqueous altered materials.\end{itemize}

\section{Acknowledgment}
The authors would like to thank Ellen Howell, Paul Lucey and Edward Cloutis for their valuable discussions and constructive suggestions. The authors also would like to thank  Zahed Wahhaj and Rachel Stevenson for reading and commenting on the manuscript. We especially thank the referee Josh Emery for his careful review and constructive comments. BY was visiting astronomer at the Infrared Telescope Facility, which is operated by the University of Hawaii under Cooperative Agreement no. NNX-08AE38A with the National Aeronautics and Space Administration, Science Mission Directorate, Planetary Astronomy Program. BY was supported by the National Aeronautics and Space Administration through the NASA Astrobiology Institute under Cooperative Agreement No. NNA08DA77A issued through the Office of Space Science and by a grant to David Jewitt from the NASA Origins program.

\clearpage

\begin{deluxetable}{lllllllll}\tablewidth{5.8in}
\tabletypesize{\scriptsize}
\tablecaption{ Observational Parameters for Trojan Asteroids
  \label{obstable}}
\tablecolumns{9} \tablehead{ \colhead{Object}    &
\colhead{ UT Date }   &  \colhead{ V }    &   \colhead{ $\alpha$} &
\colhead{r}  & \colhead{$\Delta$} & \colhead{Exp.time} & \colhead{Airmass} &
\colhead{Standard}\\
\colhead{}&\colhead{}&\colhead{mag}&\colhead{deg}&\colhead{(AU)}&\colhead{(AU)}&\colhead{(s)}&\colhead{}&\colhead{}}
\startdata
(884) Priamus & 2007 Apr. 06 &  15.75 & 3.34  & 5.093 & 4.129 &  120 x 12 &  1.15 - 1.16 &  SA1021081 \\
(884) Priamus & 2007 Apr. 07 &  15.76 & 3.50 & 5.092 & 4.132 &  120 x 10 &  1.21 - 1.26 &  HD94270  \\
(1172) Aneas & 2007 Apr. 06 &  15.43 & 3.21  & 5.275 & 4.311 &  120 x 10 &  1.31 - 1.31 &  HD115642  \\
(1172) Aneas & 2007 Apr. 07 &  15.44 & 3.25  & 5.274 & 4.311 &  120 x 8 &  1.33 - 1.35 &  HD115642  \\
(1208) Troilus & 2007 Apr. 07 &  16.43 & 3.67  & 5.617 & 4.671 &  120 x 20 &  1.34 - 1.63 & HD124019  \\  
(1208) Troilus & 2007 Apr. 07 &  16.43 & 3.60  & 5.616 & 4.668 &  120 x 20 &  1.20 - 1.44 & HD124019  \\  
(2207) Antenor & 2007 Apr. 06 &  15.71 & 1.31  & 5.140 & 4.145 &  120 x 10 &  1.19 - 1.25 &  SA105-56  \\
(2207) Antenor & 2007 Apr. 07 &  15.70 & 1.20  & 5.140 & 4.144 &  120 x 10 &  1.13 - 1.16 &  SA105-56  \\
(2223) Sarpedon & 2007 Apr. 06 &  16.38 & 2.74  & 5.168 & 4.192 &  120 x 20 &  1.23 - 1.26 &  HD115642  \\
(2223) Sarpedon & 2007 Apr. 07 &  16.36 & 2.54  & 5.168 & 4.188 &  120 x 20 &  1.23 - 1.25 &  HD115642  \\  
(2241) Alcathous & 2007 Apr. 06 &  15.56 & 6.06  & 4.879 & 3.994 &  120 x 12 &  1.25 - 1.29 &  HD94562  \\  
(2241) Alcathous & 2007 Apr. 07 &  15.57 & 6.21  & 4.879 & 4.000 &  120 x 10 &  1.37 - 1.45 &  HD94562  \\  
(2357) Phereclos & 2007 Apr. 06 &  15.55 & 0.19  & 5.057 & 4.056 &  120 x 10 &  1.17 - 1.22 &  SA105-56  \\
(3451) Mentor & 2007 Apr. 06 &  15.03 & 1.80  & 5.205 & 4.215 &  120 x  10 &  1.32 - 1.43 & SA105-56  \\
(3451) Mentor & 2007 Apr. 07 &  15.03 & 1.81  & 5.205 & 4.214 &  120 x  11 &  1.19 - 1.28 & SA105-56  \\
(3548) Eurybates & 2007 Aug. 06 &  16.63 & 5.01  & 5.175 & 4.246 &  120 x  20 &  1.31 - 1.51 & HD221495  \\
(3548) Eurybates & 2007 Aug. 07&  16.62 & 4.83  & 5.174 & 4.239 &  120 x  16 &  1.25 - 1.26 & HD213199 \\
13862 (1999 XT160) & 2007 Aug. 06 &  17.91 & 0.06  & 5.303 & 4.288 &  120 x  18 &  1.34 - 1.59 & HD221495  \\
18060 (1999 XJ156)& 2007 Aug. 06 &  16.63 & 2.40  & 5.039 & 4.042 &  120 x  18 &  1.16 - 1.25 & HD198259  \\
18060 (1999 XJ156) & 2007 Aug. 07&  16.62 & 2.56  & 5.038 & 4.045 &  120 x  24 &  1.16 - 1.23 &  HD207079 \\

\enddata
\end{deluxetable}

\clearpage

\begin{deluxetable}{lcc}\tablewidth{4.0in}
\tablecaption{Strength of the 1-$\mu$m Band \label{obstable}}
\tablecolumns{3} \tablehead{ \colhead{Object}    &
\colhead{1-$\mu$m Band Depth}   &  \colhead{1-$\mu$m Band Depth} \\
\colhead{}&\colhead{(Howell 1995)}&\colhead{(This Work)}}
\startdata
(884) Priamus & 4.3 \% &  $<$1 \% \\
(1172) Aneas &  14.1 \% &$<$1 \% \\
(1208) Troilus  & 18.9 \% &$<$1 \% \\
(2207) Antenor &  5.1 \% &$<$1 \% \\
(2223) Sarpedon & 28.9 \% &$<$1 \% \\
(2241) Alcathous &  2.7 \% &$<$1 \% \\
(2357) Phereclos & 4.1 \% &$<$1 \% \\
\enddata
\label{abs_lim}
\end{deluxetable}

\clearpage

\begin{thebibliography}{58}
\expandafter\ifx\csname natexlab\endcsname\relax\def\natexlab#1{#1}\fi

\bibitem[{Aines \& Rossman(1984)}]{aines:1984}
Aines, R.~D., \& Rossman, G.~R. 1984, Journal of Geophysical Research, 89, 4059

\bibitem[{Beaug{\'e} \& Roig(2001)}]{beauge:2001}
Beaug{\'e}, C., \& Roig, F. 2001, Icarus, 153, 391

\bibitem[{Bell(1989)}]{bell:1989}
Bell, J.~F. 1989, Icarus, 78, 426

\bibitem[{Bendjoya {et~al.}(2004)Bendjoya, Cellino, di~Martino, \&
  Saba}]{bendjoya:2004}
Bendjoya, P., Cellino, A., di~Martino, M., \& Saba, L. 2004, Icarus, 168, 374

\bibitem[Binzel et al.(2004)]{binzel:2004} Binzel, R.~P., Rivkin, 
A.~S., Stuart, J.~S., Harris, A.~W., Bus, S.~J., 
\& Burbine, T.~H.\ 2004, \icarus, 170, 259 

\bibitem[{Bradley {et~al.}(1992)Bradley, Humecki, \& Germani}]{bradley:1992}
Bradley, J.~P., Humecki, H.~J., \& Germani, M.~S. 1992, \apj,
  394, 643

\bibitem[Brearley(2006)]{brearley:2006} Brearley, A.J., 2006. The action of water. In: Lauretta, D., McSween, H.Y., Leshin, L. 
(Eds.), Meterorites and The Early Solar System II. University of Arizona Press, 
Tucson, pp. 587-624. 

\bibitem[Brown et al.(2000)]{brown:2000} Brown, P.~G., et al.\ 
2000, Science, 290, 320 

\bibitem[Brunetto et al.(2006)]{brunetto:2006} Brunetto, R., Barucci, 
M.~A., Dotto, E., \& Strazzulla, G.\ 2006, \apj, 644, 646 

\bibitem[{Burns(1981)}]{burns:1981}
Burns, R.~G. 1981, Annual Review of Earth and Planetary Sciences, 9, 345 

\bibitem[{Bus \& Binzel(2002)}]{bus:2002}
Bus, S.~J., \& Binzel, R.~P. 2002, Icarus, 158, 146

\bibitem[Chapman(1996)]{chapman:1996} Chapman, C.~R.\ 1996, Meteoritics and Planetary Science, 31, 699 

\bibitem[{Chiang \& Lithwick(2005)}]{chiang:2005}
Chiang, E.~I., \& Lithwick, Y. 2005, \apj, 628, 520

\bibitem[Cloutis et al.(1990)]{clo90} Cloutis, E.~A., Gaffey, 
M.~J., Smith, D.~G.~W., \& Lambert, R.~S.~J.\ 1990, \jgr, 95, 281 

\bibitem[{Crovisier {et~al.}(1997)Crovisier, Leech, Bockelee-Morvan, Brooke,
  Hanner, Altieri, Keller, \& Lellouch}]{crovisier:1997}
Crovisier, J., Leech, K., Bockelee-Morvan, D., Brooke, T.~Y., Hanner, M.~S.,
  Altieri, B., Keller, H.~U., \& Lellouch, E. 1997, Science, 275, 1904

\bibitem[{Cruikshank {et~al.}(2001)Cruikshank, Dalle~Ore, Roush, Geballe, Owen,
  de~Bergh, Cash, \& Hartmann}]{cruikshank:2001}
  
Cruikshank, D.~P., Dalle~Ore, C.~M., Roush, T.~L., Geballe, T.~R., Owen, T.~C.,
  de~Bergh, C., Cash, M.~D., \& Hartmann, W.~K. 2001, Icarus, 153, 348

\bibitem[{Cushing {et~al.}(2004)Cushing, Vacca, \& Rayner}]{Cushing:2004lq}
Cushing, M.~C., Vacca, W.~D., \& Rayner, J.~T. 2004, Publications of the
  Astronomical Society of the Pacific, 116, 362

\bibitem[de Luise et al.(2010)]{deluise:2010} de Luise, F., Dotto, 
E., Fornasier, S., Barucci, M.~A., Pinilla-Alonso, N., Perna, D., 
\& Marzari, F.\ 2010, \icarus, 209, 586 

\bibitem[{Dotto {et~al.}(2006)Dotto, Fornasier, Barucci, Licandro, Boehnhardt,
  Hainaut, Marzari, de~Bergh, \& de~Luise}]{dotto:2006}
Dotto, E., {et~al.} 2006, Icarus, 183, 420

 \bibitem[Dumas et al.(1998)]{dum98} Dumas, C., Owen, T., 
\& Barucci, M.~A.\ 1998, Icarus, 133, 221

\bibitem[Emery et al.(2009)]{emery:2009} Emery, J.~P., Cruikshank, 
D.~P.,  \& Burr, D.~M.\ 2009, AAS/Division for Planetary Sciences Meeting Abstracts, 41, \#32.10 

\bibitem[{Emery {et~al.}(2006)Emery, Cruikshank, \& van Cleve}]{emery:2006}
Emery, J.~P., Cruikshank, D.~P., \& van Cleve, J. 2006, Icarus, 182, 496

\bibitem[{Emery \& Brown(2004)}]{emery:2004}
Emery, J.~P., \& Brown, R.~H. 2004, Icarus, 170, 131

\bibitem[{Emery \& Brown(2003)}]{emery:2003}
Emery, J.~P., \& Brown, R.~H. 2003, Icarus, 164, 104

\bibitem[{Feierberg {et~al.}(1985)Feierberg, Lebofsky, \&
  Tholen}]{feierberg:1985}
Feierberg, M.~A., Lebofsky, L.~A., \& Tholen, D.~J. 1985, Icarus, 63, 183

\bibitem[{Fern\`andez {et~al.}(2003)Fern\`andez, Sheppard, \&
  Jewitt}]{fernandez:2003}
Fern\`andez, Y.~R., Sheppard, S.~S., \& Jewitt, D.~C. 2003, Astronomical
  Journal, 126, 1563

\bibitem[{Fleming \& Hamilton(2000)}]{fleming:2000}
Fleming, H.~J., \& Hamilton, D.~P. 2000, Icarus, 148, 479

\bibitem[{Fornasier {et~al.}(2007)Fornasier, Dotto, Hainaut, Marzari,
  Boehnhardt, de~Luise, \& Barucci}]{fornasier:2007}
Fornasier, S., Dotto, E., Hainaut, O., Marzari, F., Boehnhardt, H., de~Luise,
  F., \& Barucci, M.~A. 2007, Icarus, 190, 622


\bibitem[{Gaffey {et~al.}(1993)Gaffey, Burbine, \& Binzel}]{gaffey:1993}
Gaffey, M.~J., Burbine, T.~H., \& Binzel, R.~P. 1993, Meteoritics, 28, 161

\bibitem[{Gaffey(1976)}]{gaffey:1976} Gaffey, M.~J.\ 1976, \jgr, 81, 
905 

\bibitem[{Gehrz \& Ney(1992)}]{gehrz:1992}
Gehrz, R.~D., \& Ney, E.~P. 1992, Icarus, 100, 162

\bibitem[Gil-Hutton 
\& Brunini(2008)]{gil:2008} Gil-Hutton, R., \& Brunini, A.\ 2008, \icarus, 193, 567 

\bibitem[{Gradie \& Tedesco(1982)}]{gradie:1982}
Gradie, J., \& Tedesco, E. 1982, Science, 216, 1405

\bibitem[{Hanner(1999)}]{Hanner:1999}
Hanner, M.~S. 1999, Space Science Reviews, 90, 99

\bibitem[{Hardersen {et~al.}(2005)Hardersen, Gaffey, \& Abell}]{hardersen:2005}
Hardersen, P.~S., Gaffey, M.~J., \& Abell, P.~A. 2005, Icarus, 175, 141

\bibitem[Harker et al.(2007)]{Harker:2007} Harker, D.~E., Woodward, 
C.~E., Wooden, D.~H., Fisher, R.~S., 
\& Trujillo, C.~A.\ 2007, Icarus, 190, 432 

\bibitem[{Harker {et~al.}(2002)Harker, Wooden, Woodward, \&
  Lisse}]{harker:2002}
Harker, D.~E., Wooden, D.~H., Woodward, C.~E., \& Lisse, C.~M. 2002,
  \apj, 580, 579

\bibitem[{Hayward \& Hanner(1997)}]{hayward:1997}
Hayward, T.~L., \& Hanner, M.~S. 1997, Earth Moon and Planets, 78, 265

\bibitem[{Hiroi {et~al.}(2001)Hiroi, Zolensky, \& Pieters}]{hiroi:2001}
Hiroi, T., Zolensky, M.~E., \& Pieters, C.~M. 2001, Science, 293, 2234

\bibitem[{Howell(1995)}]{howell:1995}
Howell, E.~S. 1995, Ph.D. Thesis. THE UNIVERSITY OF ARIZONA.

\bibitem[Hunt 
\& Vincent(1968)]{hunt:1968} Hunt, G.~R., \& Vincent, R.~K.\ 1968, \jgr, 73, 6039 

\bibitem[{Huss {et~al.}(2003)Huss, Meshik, Smith, \&
  Hohenberg}]{huss:2003}
Huss, G.~R., Meshik, A.~P., Smith, J.~B., \& Hohenberg, C.~M. 2003,
  Geochimica et Cosmochimica Acta, 67, 4823

\bibitem[{Ivezic {et~al.}(2001)Ivezic, Tabachnik, Rafikov, Lupton, Quinn,
  Hammergren, Eyer, Chu, Armstrong, Fan, Finlator, Geballe, Gunn, Hennessy,
  Knapp, Leggett, Munn, Pier, Rockosi, Schneider, Strauss, Yanny, Brinkmann,
  Csabai, Hindsley, Kent, Lamb, Margon, McKay, Smith, Waddel, York, \& the
  SDSS~Collaboration}]{ivezic:2001}
Ivezic, Z., {et~al.} 2001, \aj, 122, 2749

\bibitem[{Jewitt(2002)}]{jewitt:2002}
Jewitt, D.~C. 2002, \aj, 123, 1039

\bibitem[{Jewitt \& Luu(1990)}]{jewitt:1990}
Jewitt, D.~C., \& Luu, J.~X. 1990, \aj, 100, 933

\bibitem[Jones et al.(1990)]{jon90} Jones, T.~D., Lebofsky, 
L.~A., Lewis, J.~S., \& Marley, M.~S.\ 1990, Icarus, 88, 172 

\bibitem[Kimura et 
al.(2006)]{2006A&A...449.1243K} Kimura, H., Kolokolova, L., \& Mann, I.\ 2006, \aap, 449, 1243 


\bibitem[{Kimura {et~al.}(2008)Kimura, Chigai, \& Yamamoto}]{kimura:2008}
Kimura, H., Chigai, T., \& Yamamoto, T. 2008, Astronomy and Astrophysics, 482,
  305

\bibitem[{King \& Ridley(1987)}]{king:1987}
King, T. V.~V., \& Ridley, W.~I. 1987, Journal of Geophysical Research, 92,
  11457

\bibitem[Lazzarin et al.(2006)]{lazzarin:2006} Lazzarin, M., Marchi, 
S., Moroz, L.~V., Brunetto, R., Magrin, S., Paolicchi, P., 
\& Strazzulla, G.\ 2006, \apjl, 647, L179 

\bibitem[{Lucey \& Noble(2008)}]{lucey:2008}
Lucey, P.~G., \& Noble, S.~K. 2008, Icarus, 197, 348

\bibitem[Luu et al.(1994)]{luu94} Luu, J., Jewitt, D., 
\& Cloutis, E.\ 1994, Icarus, 109, 133

\bibitem[{Marzari \& Scholl(1998)}]{marzari:1998}
Marzari, F., \& Scholl, H. 1998, Icarus, 131, 41

\bibitem[Moersch \& Christensen(1995)]{moersch:1995} Moersch, J.~E., \& Christensen, P.~R.\ 1995, \jgr, 100, 7465 

\bibitem[{Morbidelli {et~al.}(2005)Morbidelli, Levison, Tsiganis, \&
  Gomes}]{morbidelli:2005}
Morbidelli, A., Levison, H.~F., Tsiganis, K., \& Gomes, R. 2005, Nature, 435,
  462

\bibitem[{Moroz {et~al.}(2004)Moroz, Baratta, Strazzulla, Starukhina, Dotto,
  Barucci, Arnold, \& Distefano}]{moroz:2004}
Moroz, L., Baratta, G., Strazzulla, G., Starukhina, L., Dotto, E., Barucci,
  M.~A., Arnold, G., \& Distefano, E. 2004, Icarus, 170, 214
  
  \bibitem[Moroz et al.(2004b)]{moroz:2004b} Moroz, L.~V., Hiroi, T., 
Shingareva, T.~V., Basilevsky, A.~T., Fisenko, A.~V., Semjonova, L.~F., 
\& Pieters, C.~M.\ 2004, Lunar and Planetary Institute Science Conference Abstracts, 35, 1279 

\bibitem[Moroz et al.(1998)]{moroz:1998} Moroz, L.~V., Arnold, G., 
Korochantsev, A.~V., \& Wasch, R.\ 1998, \icarus, 134, 253 

\bibitem[{Mukai \& Koike(1990)}]{mukai:1990}
Mukai, T., \& Koike, C. 1990, Icarus, 87, 180

\bibitem[Mustard 
\& Hays(1997)]{mustard:1997} Mustard, J.~F., \& Hays, J.~E.\ 1997, \icarus, 125, 145 


\bibitem[Nesvorn{\'y} et al.(2005)]{nesvorny:2005} Nesvorn{\'y}, D., 
Jedicke, R., Whiteley, R.~J., 
\& Ivezi{\'c}, {\v Z}.\ 2005, \icarus, 173, 132 

\bibitem[{Ney(1977)}]{ney:1977}
Ney, E.~P. 1977, Science, 195, 541


\bibitem[{Rayner {et~al.}(2003)Rayner, Toomey, Onaka, Denault, Stahlberger,
  Vacca, Cushing, \& Wang}]{rayner:2003}
Rayner, J.~T., Toomey, D.~W., Onaka, P.~M., Denault, A.~J., Stahlberger, W.~E.,
  Vacca, W.~D., Cushing, M.~C., \& Wang, S. 2003, \pasp, 115, 362

\bibitem[{Rivkin {et~al.}(2002)Rivkin, Howell, Vilas, \&
  Lebofsky}]{rivkin:2002}
Rivkin, A.~S., Howell, E.~S., Vilas, F., \& Lebofsky, L.~A. 2002, Asteroids
  III, W. F. Bottke Jr., A. Cellino, P. Paolicchi, and R. P. Binzel (eds), University of Arizona Press, Tucson, 235

\bibitem[Roig et al.(2008)]{roig:2008} Roig, F., Ribeiro, A.~O., \& Gil-Hutton, R.\ 2008, \aap, 483, 911 

\bibitem[{Rose(1979)}]{rose:1979}
Rose, L.~A. 1979, Astrophysics and Space Science, 65, 47


\bibitem[Salisbury 
\& Wald(1992)]{salisbury:1992} Salisbury, J.~W., \& Wald, A.\ 1992, \icarus, 96, 121 

\bibitem[Shingareva et al.(2004)]{shingareva:2004} Shingareva, T.~V., 
Basilevsky, A.~T., Fisenko, A.~V., Semjonova, L.~F., 
\& Korotaeva, N.~N.\ 2004, Lunar and Planetary Institute Science Conference Abstracts, 35, 1137 

\bibitem[{Stansberry {et~al.}(2004)Stansberry, Van~Cleve, Reach, Cruikshank,
  Emery, Fernandez, Meadows, Su, Misselt, Rieke, Young, Werner, Engelbracht,
  Gordon, Hines, Kelly, Morrison, \& Muzerolle}]{stansberry:2004}
Stansberry, J.~A., {et~al.} 2004, Astrophysical Journal Supplement Series, 154,
  463

\bibitem[{Stephens \& Rusell(1979)}]{stephens:1979}
Stephens, J.~R., \& Rusell, R.~W. 1979, \apj, 228, 780

\bibitem[{Sunshine \& Pieters(1998)}]{sunshine:1998}
Sunshine, J.~M., \& Pieters, C. 1998, Journal of Geophysical Research, 103,
  13675

\bibitem[{Tedesco {et~al.}(2002)Tedesco, Noah, Noah, \& Price}]{tedesco:2002}
Tedesco, E.~F., Noah, P.~V., Noah, M., \& Price, S.~D. 2002, \aj, 123, 1056

\bibitem[{Tholen \& Barucci(1989)}]{tholen:1989}
Tholen, D.~J., \& Barucci, M.~A. 1989, in Asteroids II, ed. R. P. Binzel, T. Gehrels, \& M. S. Matthews (Tucson, AZ: Univ. of Arizona Press), 298 

\bibitem[{Wooden {et~al.}(1999)Wooden, Harker, Woodward, Butner, Koike,
  Witteborn, \& McMurtry}]{wooden:1999}
Wooden, D.~H., Harker, D.~E., Woodward, C.~E., Butner, H.~M., Koike, C.,
  Witteborn, F.~C., \& McMurtry, C.~W. 1999, \apj, 517, 1034

\bibitem[{Yang \& Jewitt(2007)}]{yang:2007}
Yang, B., \& Jewitt, D. 2007, \aj, 134, 223

\bibitem[Szab{\'o} et al.(2007)]{szab:2007} Szab{\'o}, G.~M., 
Ivezi{\'c}, {\v Z}., Juri{\'c}, M., \& Lupton, R.\ 2007, \mnras, 377, 1393 

\bibitem[Zellner {et al.}(1985)]{zellner:1985} Zellner, B., Tholen, 
D.~J., \& Tedesco, E.~F.\ 1985, \icarus, 61, 355 

\bibitem[{Zellner(1975)}]{zellner:1975}
Zellner, B. 1975, \apj, 198, L45


\end{thebibliography}
  \end{document}